\documentclass[conference]{IEEEtran}
\IEEEoverridecommandlockouts
\usepackage{cite}
\usepackage{amsmath,amssymb,amsfonts}
\usepackage{amsthm}
\usepackage{algorithmic}
\usepackage{graphicx}
\usepackage{textcomp}
\usepackage{xcolor}
\usepackage{pifont}
\usepackage{caption}
\usepackage{subcaption}
\usepackage{booktabs}
\usepackage{hyperref}

\def\BibTeX{{\rm B\kern-.05em{\sc i\kern-.025em b}\kern-.08em
    T\kern-.1667em\lower.7ex\hbox{E}\kern-.125emX}}
\begin{document}

\newcommand{\mynote}[1]{%
               \fbox{\color{red}\bfseries\sffamily\scriptsize#1}}
\newcommand{\ljw}[1]{\mynote{Liao:}{\color{red}#1}}

\newtheorem{theorem}{Theorem}
\newtheorem{lemma}{Lemma}
\newtheorem{definition}{Definition}

\title{
Learned Data Compression: Challenges and Opportunities for the Future
\thanks{Correspondence to Dr.~Qiyu Liu (Email: \href{mailto:qyliu.cs@gmail.com}{qyliu.cs@gmail.com}). }
}

\author{
\IEEEauthorblockN{Qiyu Liu\IEEEauthorrefmark{1}, Siyuan Han\IEEEauthorrefmark{2}, Jianwei Liao\IEEEauthorrefmark{1}, Jin Li\IEEEauthorrefmark{3}, Jingshu Peng\IEEEauthorrefmark{4}, Jun Du\IEEEauthorrefmark{5}, Lei Chen\IEEEauthorrefmark{2}\IEEEauthorrefmark{6}}

\IEEEauthorblockA{
\IEEEauthorrefmark{1}Southwest University, 
\IEEEauthorrefmark{2}HKUST, 
\IEEEauthorrefmark{3}Harvard University, 
\IEEEauthorrefmark{4}ByteDanace, 
\IEEEauthorrefmark{5}Hong Kong University,
\IEEEauthorrefmark{6}HKUST (GZ)\\
\{qyliu.cs, liaotoad\}@gmail.com, 
\{shanaj, leichen\}@cse.ust.hk,
jinli@g.harvard.edu,\\
jingshu.peng@bytedance.com,
jundu2024@connect.hku.hk
}
}

\maketitle

\begin{abstract}
Compressing integer keys is a fundamental operation among multiple communities, such as database management (DB), information retrieval (IR), and high-performance computing (HPC). 
Recent advances in \emph{learned indexes} have inspired the development of \emph{learned compressors}, which leverage simple yet compact machine learning (ML) models to compress large-scale sorted keys. 
The core idea behind learned compressors is to \emph{losslessly} encode sorted keys by approximating them with \emph{error-bounded} ML models (e.g., piecewise linear functions) and using a \emph{residual array} to guarantee accurate key reconstruction. 

While the concept of learned compressors remains in its early stages of exploration, our benchmark results demonstrate that an SIMD-optimized learned compressor can significantly outperform state-of-the-art CPU-based compressors. Drawing on our preliminary experiments, this vision paper explores the potential of learned data compression to enhance critical areas in DBMS and related domains. 
Furthermore, we outline the key technical challenges that existing systems must address when integrating this emerging methodology. 

\end{abstract}


\section{Introduction}\label{sec:introduction}
\begin{figure}[t]
    \centering
    \includegraphics[width=0.45\textwidth]{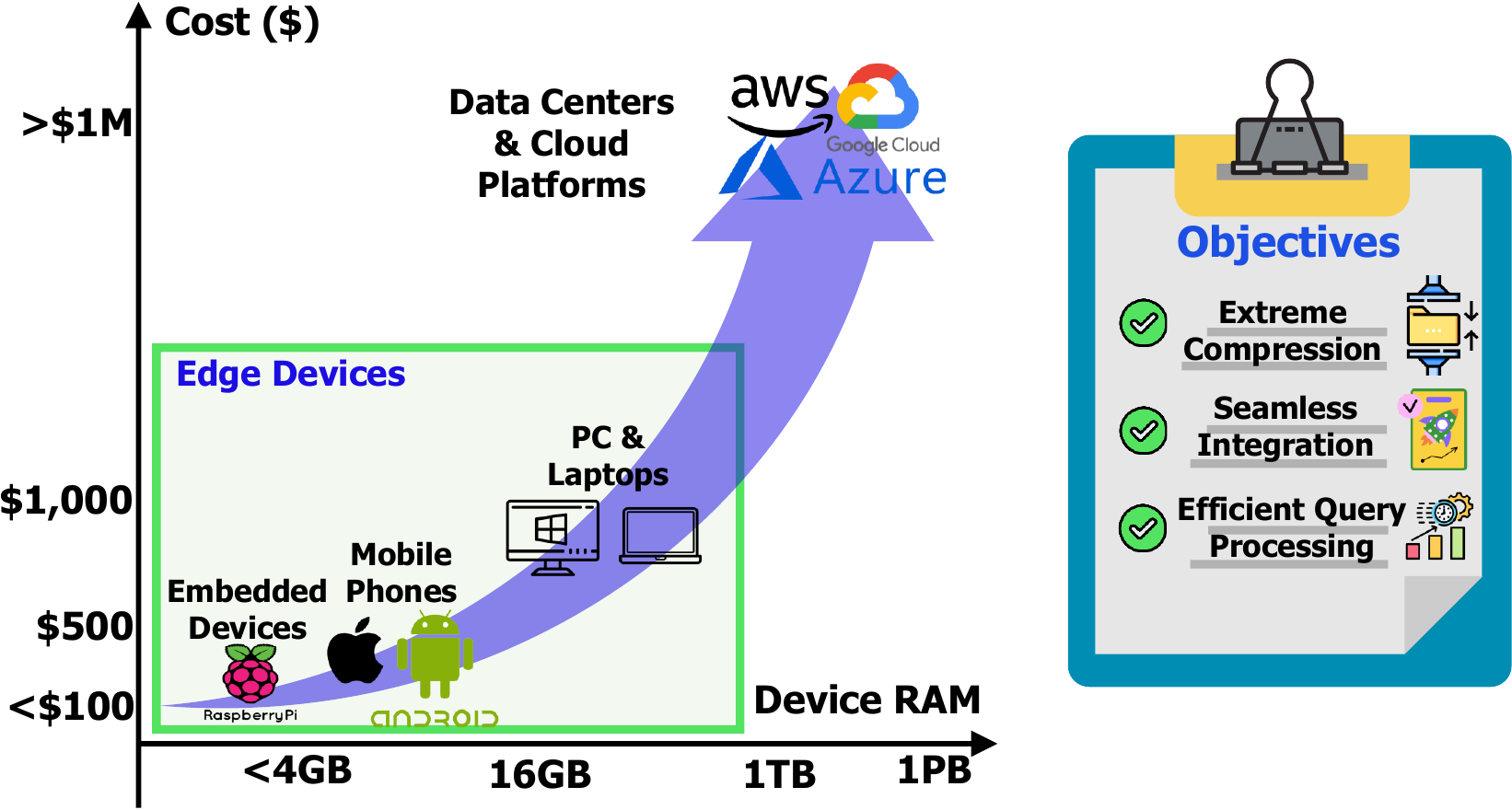}
    \caption{(left) Illustration of the disproportionate growth between the device memory capacities and their associated costs. (right) Key design objectives for a compression algorithm: \ding{182}~high compression ratio, \ding{183}~seamless integration with existing systems, and \ding{184}~efficient query processing, ideally enabling direct query processing on the compressed data. }
    \label{fig:memory_motivation}
    \vspace{-2ex}
\end{figure}

A defining feature of modern data-driven systems is the unprecedented scale of data they manage and store~\cite{marx2013big,trelles2011big}. 
As data volumes continue to grow exponentially, the demand for storage in data analytics is increasing rapidly. 
As illustrated in Figure~\ref{fig:memory_motivation}, unfortunately, compared to the explosive growth of data, advancements in storage (in terms of cost per GiB) have been far less significant. 
Consequently, data compression has become a critical component in big data management and analytics, especially within resource-constrained environments~\cite{satyanarayanan2011mobile,lin2019computation}. 
An effective data compression algorithm can significantly reduce I/O and communication overhead, thereby improving system performance.

The design philosophies of compression algorithms vary significantly across different data types, such as integers, floats, and strings. 
In this paper, we focus on \emph{lossless sorted integer key compression}, which is widely used in database systems and information retrieval applications, such as key-value (KV) stores~\cite{rocksdb,luo2020lsm}, distributed systems~\cite{ozsu1999principles,geambasu2010comet}, inverted indexes~\cite{pibiri2020techniques}, and vector databases~\cite{pan2024survey}. 
Formally, the problem of sorted integer compression is defined as follows.

\begin{definition}[Lossless Integer Compression]
    Given a sorted list of $N$ integer keys $\mathcal{K}=\{k_1,k_2,\cdots,k_N\}$, where $k_i\in\mathbb{Z}$ is from a universe of $U=\{a, a+1, \cdots, b\}$. 
    The problem of lossless integer compression is to find a compact encoding scheme $\mathcal{K}^c$ together with a decoding function $\mathsf{Dec}(\cdot)$, such that $\mathsf{Dec}(\mathcal{K}^c)[i]=\mathcal{K}[i]$ for $\forall i=1, \cdots, N$.  
    The compression ratio is defined as $r = \mathsf{size}(\mathcal{K}^c)/\mathsf{size}(\mathcal{K})$. 
\end{definition}

As a classic problem, integer compression has been extensively studied by the database, information retrieval (IR), and high-performance computing (HPC) communities for decades, with related algorithms widely deployed in practical systems. 
Existing solutions can be majorly classified into three families: 
\ding{182}~\textbf{Generic Integer Compression}, such as $\delta$-code~\cite{elias1975universal}, Golomb~\cite{golomb1966run,teuhola1978compression}, Rice~\cite{rice1971adaptive}, and Variable-Byte~\cite{stepanov2011simd,dean2009challenges}; 
\ding{183}~\textbf{Ordered List Compression}: such as P4Delta~\cite{zukowski2006super}, OptP4Delta~\cite{yan2009inverted}, Elias-Fano Index~\cite{elias1974efficient,ottaviano2014partitioned,vigna2013quasi}, and Binary Interpolative Code (BIC)~\cite{moffat1996exploiting,moffat2000binary}; 
and \ding{184}~\textbf{Entropy Encoding}: such as Huffman encoding~\cite{huffman1952method}, Asymmetric Numeral System (ANS)~\cite{duda2013asymmetric}, and Arithmetic coding~\cite{moffat1998arithmetic}. 
A recent experimental survey~\cite{pibiri2020techniques} systematically summarizes and evaluates the mainstream approaches to sorted integer list compression, from the perspective of inverted index compression in full-text search engines.

\begin{figure*}[t]
    \centering
    \includegraphics[width=0.8\textwidth]{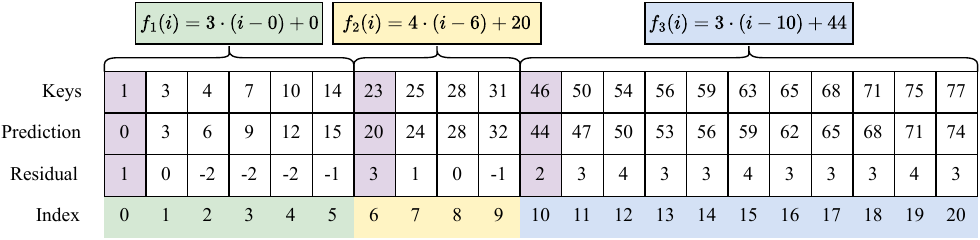}
    \caption{A toy example of learned integer compressor based on an error-bounded PLA model.  \ding{182}~\textbf{Encoding Stage}: By setting $\epsilon=4$, a PLA model with 3 line segments can be fitted such that each residual $|\delta_i|=|\mathcal{K}[i]-\lfloor f(i) \rfloor|\leq\epsilon$. 
    Segments $f_1,f_2,f_3$ and all residuals, $\Delta=\{1, 0, \cdots, 3\}$, are materialized as a compressed version to the original key set $\{1, 3, \cdots, 77\}$. 
    Encoding each residual requires $\lceil\log_2(2\epsilon+1)\rceil=4$ bits. \ding{183}~\textbf{Decoding Stage}: For each index $i\in[0, 5]$, the original key can be losslessly recovered by $\mathcal{K}[i]=\lfloor f_1(i)\rfloor+\Delta[i]$. 
    Similarly, $\mathcal{K}[i]=\lfloor f_2(i)\rfloor+\Delta[i]$ for $i\in[6, 9]$, and $\mathcal{K}[i]=\lfloor f_3(i)\rfloor+\Delta[i]$ for $i\in[10, 20]$.
    }
    \label{fig:compression_eg}
    \vspace{-2ex}
\end{figure*}

Inspired by the seminal work on \emph{learned index}~\cite{kraska2018case}, recent studies have begun exploring the use of simple machine learning (ML) models for data compression, leading to a new paradigm named \emph{learned compression}~\cite{boffa2022learned,boffa2021learned}. 
Given an integer list $\mathcal{K}$, learned compressors aim to directly fit a projection function $f:\mathcal{I}\mapsto\mathcal{K}$ with a controllable error constraint $\epsilon$, where $\mathcal{I}=\{1,2,\cdots,|\mathcal{K}|\}$ is the index set. 
With an error-bounded $f$, the sorted list $\mathcal{K}$ can be encoded as $\mathcal{K}^{c}=(f, \Delta)$, where $\Delta$ is an array of \emph{low-bit} residuals ($\lceil\log_2(2\epsilon+1)\rceil$ bits) between the model's predictions and the actual keys, i.e., $\Delta[i]=\mathcal{K}[i]-\lfloor f(i)\rfloor$. 
Given $\mathcal{K}^c$, the decompression, which is \emph{lossless}, is easy as $\mathcal{K}[i]=\lfloor f(i)\rfloor+\Delta[i]$.

Intuitively, the compression efficacy hinges on the ability of a compact model $f$ to approximate this mapping with low maximum error, requiring a careful balance between model size and residual size. 
Boffa et~al.~propose \textsf{la-vector}~\cite{boffa2021learned,boffa2022learned} as the first learned compression algorithm, building on their earlier work PGM-Index~\cite{ferragina2020pgm,ferragina2020learned}. 
Similar to the PGM-Index, \textsf{la-vector} employs the error-bounded piecewise linear approximation ($\epsilon$-PLA) model to fit the mapping $f:\mathcal{I}\mapsto \mathcal{K}$, and Figure~\ref{fig:compression_eg} illustrates a toy example of PLA-based learned compression.  

Unlike learned indexes\footnote{As of December 1, 2024, a search for the term ``learned index'' on DBLP returned 205 matches. Ref: \url{https://dblp.org/search?q=learned+index}}, which have attracted significant attention from both academia and industry since the seminal work by Kraska et~al.~\cite{kraska2018case}, the concept of learned compression, introduced in 2021~\cite{boffa2021learned}, has yet to gain widespread adoption or undergo thorough exploration. 
The core reason is that, compared to conventional integer compressors like OptP4Delta~\cite{yan2009inverted} and BIC~\cite{moffat1996exploiting}, current learned compressors like \textsf{la-vector} are not fully optimized and cannot show dominant performance in practice, in terms of both compression ratio and decompression throughput. 
Moreover, conventional methods have been deeply integrated into DBMS and full-text search engines for decades, such as PostgreSQL~\cite{postgres}, Oracle~\cite{oracle}, Solr~\cite{solr}, and Elasticsearch~\cite{elasticsearch}. As a result, practitioners have shown limited interest in transitioning to the relatively immature learned data compression techniques.

However, as we will demonstrate in Section~\ref{sec:pre}, an optimized implementation of learned compression that fully leverages the SIMD parallelism available in modern architectures can achieve a comparable compression ratio while delivering SOTA decompression throughput on CPUs. 
Specifically, learned compressor with SIMD achieves a throughput of \textbf{6.535} GiB/sec, which is \textbf{1.68\texttimes} faster than OptP4Delta~\cite{yan2009inverted} and \textbf{18.25\texttimes} faster than BIC~\cite{moffat1996exploiting}, on a web-scale inverted index compression benchmark~\cite{ccnews2020}. 
These preliminary results suggest that, though requiring substantial optimization, the learned data compression methodology has the potential to challenge and redefine the role of widely used data compression algorithms in modern data-intensive systems.

We emphasize the potential advantages of learned data compression in three key aspects.  
\ding{182}~\textbf{High Performance}: a well-optimized SIMD-based implementation can outperform the existing CPU-based compressors. 
\ding{183}~\textbf{Flexibility}: existing systems can obtain a seamless improvement by replacing original compression algorithms with learned compressors. 
\ding{184}~\textbf{Potential Query Support}: learned compressors can be extended to efficiently process a wide range of queries, such as list intersection, union, and approximate quantiles. 

Building on these advantages, we claim that learned data compression can serve as a new foundation for systems requiring high-performance integer compression. 
In this vision paper, we comprehensively explore the potential applications and technical challenges of learned data compression in addressing conventional DB and IR problems, such as full-text search engines~\cite{solr,pibiri2020techniques}, KV store compression~\cite{geambasu2010comet}, approximate query processing~\cite{chaudhuri2017approximate,cormode2011sketch}, and vector databases~\cite{pan2024survey}.


The remainder of this paper is structured as follows. 
Section~\ref{sec:pre} provides an overview of the preliminaries and presents the motivation experiments. 
Section~\ref{sec:applications} explores how and why existing DBMS and IR systems can benefit from integrating learned data compression. 
In Section~\ref{sec:challenge}, we discuss the technical challenges to be addressed to facilitate the deployment of this promising approach in real-world systems. 

\begin{figure*}
     \centering
     \begin{subfigure}[b]{0.5\textwidth}
         \centering
{\small         \begin{tabular}{c|cccc}
\textbf{\textsf{Method}} & \textbf{\begin{tabular}[c]{@{}c@{}}\textsf{Total Size}\\ \textsf{GiB}\end{tabular}} & \textbf{\begin{tabular}[c]{@{}c@{}}\textsf{Avg.~Size}\\ \textsf{bits/int}\end{tabular}} & \textbf{\begin{tabular}[c]{@{}c@{}}\textsf{Dec.~Time}\\ \textsf{ns/int}\end{tabular}} & \textbf{\begin{tabular}[c]{@{}c@{}}\textsf{Throughput}\\ \textsf{GiB/s}\end{tabular}} \\ \hline
\textsf{bic}             & \textbf{7.472}                                                                               & \textbf{7.384}                                                                                   & 10.41                                                                                  & \multicolumn{1}{c|}{0.358}                                                            \\
\textsf{rice}            & 7.740                                                                               & 7.382                                                                                   & 4.00                                                                                   & \multicolumn{1}{c|}{0.931}                                                            \\
\textsf{simple16}        & 8.556                                                                               & 8.455                                                                                   & 2.01                                                                                   & \multicolumn{1}{c|}{1.853}                                                            \\
\textsf{optpfor}         & 8.400                                                                               & 8.301                                                                                   & 1.32                                                                                   & \multicolumn{1}{c|}{2.822}                                                            \\
\textsf{vbyte}           & 13.807                                                                              & 13.743                                                                                  & 1.33                                                                                   & \multicolumn{1}{c|}{2.801}                                                            \\
\textsf{opt-vbyte}       & 9.090                                                                               & 8.983                                                                                   & 0.96                                                                                   & \multicolumn{1}{c|}{3.881}                                                            \\
\textsf{slicing}         & 9.545                                                                               & 9.432                                                                                   & 0.89                                                                                   & \multicolumn{1}{c|}{4.186}                                                            \\
\textsf{qmx}             & 9.817                                                                               & 9.701                                                                                   & 0.74                                                                                   & \multicolumn{1}{c|}{5.034}                                                            \\ \cline{2-5} 
\textsf{lc}              & 8.947                                                                               & 8.841                                                                                   & 0.81                                                                                   & \multicolumn{1}{c|}{4.599}                                                            \\
\textsf{lc-simd}         & 8.947                                                                               & 8.841                                                                                   & \textbf{0.57}                                                                                   & \multicolumn{1}{c|}{\textbf{6.535}}                                                            \\ \cline{2-5} 
\end{tabular}}
         \caption{Results on dataset \textsf{CCNews}~\cite{ccnews2020}.}\label{subfig:ccn_results}
     \end{subfigure}
     \hfill
     \begin{subfigure}[b]{0.45\textwidth}
         \centering
         \includegraphics[width=\textwidth]{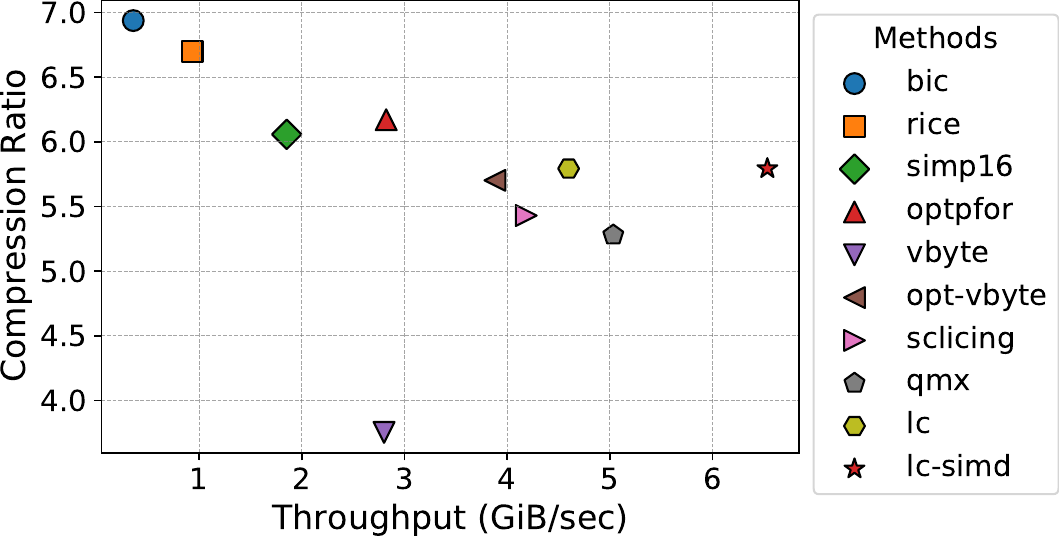}
         \caption{Compression ratio v.s.~decompression throughput.}\label{subfig:vis_methods}
     \end{subfigure}
        \caption{Preliminary benchmark results for the original learned compressor implementation~\cite{boffa2022learned} (\textsf{lc}, a.k.a.~\textsf{la-vector}) and our SIMD-based optimization (\textsf{lc-simd}). Note we enable the compiler's auto-vectorization for \textsf{lc}. All the other baselines are chosen from a recent benchmark for inverted index compression~\cite{yan2009inverted}. All the experiments are performed on a Ubuntu machine with an Intel\copyright~Xeon\texttrademark~Gold 6430 CPU and 512 GiB DDR5 memory. }
        \label{fig:exp_result}
        \vspace{-3ex}
\end{figure*}

\section{Learned Compression Foundation}\label{sec:pre}
In this section, we introduce the foundations of learned compression methodology and report some preliminary benchmark results for motivation. 

\subsection{Learned Compression Methodology}
Intuitively, learning the mapping from the index space to the key space, i.e., $f: \mathcal{I}\mapsto\mathcal{K}$, is equivalent to learning the inverse cumulative distribution function (ICDF, a.k.a.~quantile function) of $\mathcal{K}$. 
To balance the model's expressivity and inference efficiency, existing learned compressors like \textsf{la-vector}~\cite{boffa2021learned,boffa2022learned} opt to employ simple models like piecewise linear functions to approximate $f$. 
The error-bounded piecewise linear approximation ($\epsilon$-PLA) model is defined as follows. 

\begin{definition}[$\epsilon$-PLA]\label{def:pla}
    Given a set of points in Cartesian space $\{(i,\mathcal{K}[i])\}_{i=1,\cdots,N}\subseteq\mathcal{I}\times\mathcal{K}$, an $\epsilon$-PLA is defined as a piecewise linear function of $L$ line segments,
    \begin{equation}\label{eq:pla}
         f(i)=\begin{cases}\alpha_1\cdot (i-s_1)+\beta_1&\text{ if } s_1\leq i< s_2\\
         \alpha_2\cdot (i-s_2)+\beta_2&\text{ if } s_2\leq i<s_3\\
         \quad\cdots&\qquad\cdots\\
         \alpha_{L}\cdot (i-s_L)+\beta_{L}&\text{ if } s_{L}\leq i\\\end{cases}
     \end{equation}
     such that $|\mathcal{K}[i] - \lfloor f(i)\rfloor|\leq\epsilon$ holds for $\forall i=1,\cdots,N$. 
     Each segment in $f$ is a tuple $(s_j, \alpha_j, \beta_j)$, where $s_j$ is the starting index, $\alpha_j$ is the slope, and $\beta_j$ is the intercept. 
\end{definition}

Consider a scenario where an $\epsilon$-PLA $f$, comprising $L$ line segments, is fitted for a sorted key set $\mathcal{K}$. 
The \textsf{la-vector} method losslessly encodes $\mathcal{K}$ as $\mathcal{K}^c=(f, \Delta)$ where $\Delta=\{\mathcal{K}[i]-\lfloor f(i)\rfloor \mid i=1,\cdots N\}$. 
Since $f$ is error-bounded, each residual satisfies $\Delta[i]\in[-\epsilon,+\epsilon]$. 
Thus, encoding each residual requires $\lceil\log_2(2\epsilon+1)\rceil$ bits, and the total bits can be expressed as $\mathsf{size}(\mathcal{K}^{c})=\mathsf{size}(f)+N\cdot\lceil\log_2(2\epsilon+1)\rceil$, where $\mathsf{size}(f)$ is the bits required to encode $f$. 
For uint32 keys and float32 slopes/intercepts, encoding each segment requires 12 bytes, meaning that $\mathsf{size}(f)=96L$ bits. 
To efficiently learn $f$ satisfying a given error constraint, \cite{boffa2021learned,boffa2022learned} employ an optimal online $\epsilon$-PLA fitting algorithm~\cite{ORourke81} to minimize the number of required line segments $L$ in linear time $O(N)$. 

Similar to conventional compressors, learned compressors need to tune hyper-parameters (i.e., $\epsilon$ in PLA) to achieve optimal performance. 
Intuitively, the choice of $\epsilon$ is a trade-off: 
a large $\epsilon$ reduces the required segment count but increases the bits needed per residual, while a small $\epsilon$ saves bits per residual at the cost of introducing more segments to satisfy the error constraint. 
Building on the theoretical foundations of learned indexes~\cite{ferragina2020learned,liu2024learned,ferragina2020pgm}, a recent study BitTuner~\cite{bittuner} derives the minimized total space cost of a learned compressor based on $\epsilon$-PLA as $B^{\mathsf{opt}} \approx N \cdot \left(1.721 + \left\lceil \log_2\epsilon^{\mathsf{opt}}\right\rceil\right)$, 
where the corresponding $\epsilon^{\mathsf{opt}}$ is given by $\epsilon^{\mathsf{opt}} = \sqrt{2\ln 2 \cdot\sigma^2\cdot C\cdot (K+2F)}$. 
Here, $\sigma^2$ is the variance of key gaps (i.e., $\mathcal{K}[i]-\mathcal{K}[i-1]$), $C$ is a data-dependent constant, and $K$ and $F$ denote the number of bits to encode an integer (i.e., keys) and a float (i.e., slopes and intercepts), respectively. 

\subsection{Connection with Learned Index}
Although learned indexing and learned compression appear similar, their design objectives exhibit fundamental differences. 
Intuitively, learned indexing and learned compression are \emph{``dual''} problems. 
Given a sorted key set $\mathcal{K}$, learned indexes like~\cite{kraska2018case,ferragina2020pgm,ding2020alex} model the cumulative distribution function (CDF) of $\mathcal{K}$ (i.e., the mapping of $g: \mathcal{K}\mapsto\mathcal{I}$), whereas learned compression~\cite{boffa2021learned,boffa2022learned} tries to fit the inverse CDF (i.e., the mapping of $f: \mathcal{I}\mapsto\mathcal{K}$), a.k.a.~the quantile function~\cite{serfling2009approximation}. 

The key differences between learned indexing and learned compression are threefold. 
\ding{182}~\textbf{Theoretical Foundation}: intuitively, when the data universe is large, learning the ICDF is more challenging than learning the CDF; conversely, when the universe is small, learning the CDF becomes more difficult~\cite{ferragina2020learned,boffa2022learned,bittuner}. 
\ding{183}~\textbf{Data Structure}: learned indexes like RMI~\cite{kraska2018case} and PGM-Index~\cite{ferragina2020pgm} are typically hierarchical structures optimized to reduce lookup latency. In contrast, learned compressors employ a flat structure (i.e., one level) to save space. 
\ding{184}~\textbf{Query Support}: learned indexes are designed as alternatives for conventional B+-tree indexes~\cite{graefe2011modern}, necessitating efficient support for lookup and range queries. On the other hand, learned compressors primarily focus on decompression operations (i.e., reconstructing $\mathcal{K}[i]$). Supporting additional queries, such as intersection, union, or rank, requires further algorithmic and engineering efforts.

\subsection{Preliminary Benchmark Results}\label{subsec:benchmark}
As discussed in Section~\ref{sec:introduction}, the first learned compressor \textsf{la-vector}~\cite{boffa2021learned,boffa2022learned} cannot significantly outperform conventional methods in terms of compression ratio and decompression efficiency. 
After carefully revisiting the workflow of PLA-based learned compressors, we propose SIMD-aware optimizations to fully exploit the hardware parallelism available on most modern architectures\footnote{Customer-grade CPUs like Intel\copyright~Core\texttrademark~and AMD\copyright~Ryzen\texttrademark~typically provide 256-bit SIMD registers (i.e., AVX2), while enterprise CPUs like Intel\copyright~Xeon\texttrademark~are equipped with wider 512-bit SIMD registers (i.e., AVX512).
Besides SIMD parallelism on X86 platforms, we also plan to release an ARM-based version and a GPU-accelerated version in the future.}. 
It is important to note that our implementation is not an embarrassingly parallel algorithm, as it incorporates techniques such as task decomposition, operator fusion, and memory alignment. 
The technical details of SIMD optimization can be found in~\cite{salad}. 

We evaluate our learned compressor implementation using a benchmark designed for inverted index compression~\cite{pibiri2020techniques}. 
Eight integer compressors are adopted as baselines, covering mainstream methodologies employed in practical systems. 
All the compared methods are implemented in C++ and compiled with g++ using the -O3 optimization, with auto-vectorization enabled. 
The results on a large-scale document inverted index compression task, CCNews~\cite{ccnews2020} (approx.~53 GiB of integer document IDs), are presented in Figure~\ref{fig:exp_result}. 
As seen in Figure~\ref{subfig:ccn_results}, the learned compressor with SIMD optimization (i.e., \textsf{lc-simd}) achieves a decompression throughput of \textbf{6.535 GiB/s}, which is \textbf{18.254\texttimes}, \textbf{2.316\texttimes}, \textbf{2.333\texttimes}, and \textbf{1.298\texttimes} faster than BIC~\cite{teuhola1978compression}, OptP4Delta~\cite{yan2009inverted}, Variable-Byte~\cite{stepanov2011simd}, and QMX~\cite{trotman2016vacuo}, respectively. 
Notably, most of the compared methods are also highly optimized using SIMD. 
Specifically, \textsf{lc-simd} is \textbf{1.421\texttimes} faster than \textsf{lc}, which is essentially the \textsf{la-vector}~\cite{boffa2021learned,boffa2022learned} optimized with compiler's auto-vectorization. 
Figure~\ref{subfig:vis_methods} further visualizes the trade-off between compression ratio and decompression throughput for each method, highlighting that \textsf{lc-simd} achieves comparable compression power while delivering the highest throughput among \textbf{all CPU-based} baselines. 
Results on other datasets exhibit similar trends and are excluded from this paper for brevity. 

In summary, this preliminary benchmark study reveals that with proper low-level optimizations, the learned compressor can emerge as a strong competitor, challenging the conventional compression algorithms used for decades. 
Moreover, our \textsf{lc-simd} implementation can be further enhanced by considering better memory alignment and pre-fetching~\cite{lee2012prefetching}. 
In the subsequent sections, we will discuss optimization opportunities and technical challenges to deploy this emerging technique in practical DBMS and information retrieval systems.

\section{Application Scenarios}\label{sec:applications}
In this section, we explore potential application scenarios that can benefit from the high performance of learned compression methodologies. 

\subsection{Inverted Index Compression}
The inverted index (a.k.a.~inverted lists or posting lists) in full-text search engines like Solr~\cite{solr} and Elasticsearch~\cite{elasticsearch} represents a key application of integer compression. 
Given a collection of documents, each distinct term $t$ is associated with an inverted index $I_t$, which is an ordered list of integers storing the identifiers of the documents where $t$ appears~\cite{singhal2001modern}. 
For web-scale document collections, the raw size of inverted indexes can easily scale to hundreds of gigabytes. 
As a result, compressing inverted indexes is necessary to reduce memory footprint and improve query throughput~\cite{pibiri2020techniques}. 

\begin{figure}[t]
    \centering
    \centering
     \begin{subfigure}[b]{0.4\textwidth}
         \centering
         \includegraphics[width=\textwidth]{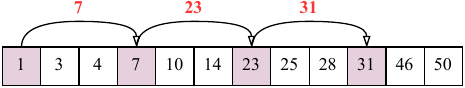}
         \caption{Illustration of skip pointers.}
         \label{subfig:skip_pointer}
     \end{subfigure}
     
     \begin{subfigure}[b]{0.4\textwidth}
         \centering
         \includegraphics[width=\textwidth]{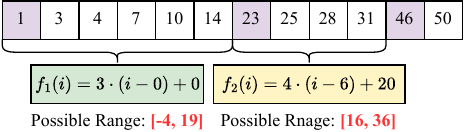}
         \caption{Illustration of segment-based pruning.}\label{subfig:segment_skip}
     \end{subfigure}
     
    \caption{(a) Skip pointers in conventional inverted index designs. (b) Natural pruning support in learned compressors. The possible key range covered by a line segment is known due to the error-bounded nature of an $\epsilon$-PLA.}
    \label{fig:skipping}
    \vspace{-3ex}
\end{figure}

List intersection (AND queries) and list union (OR queries) are fundamental operations for learned compressors to ensure compatibility with existing full-text search engines. 
Fortunately, learned compressors inherently support these queries efficiently by utilizing segment-level pruning. 
As illustrated in Figure~\ref{subfig:skip_pointer}, traditional compressors such as P4Delta~\cite{zukowski2006super} and Elias-Fano Index~\cite{vigna2013quasi} often implement auxiliary shortcuts called ``skip pointers'' that bypass sections of inverted lists that do not appear in the final query results. 
While skip pointers improve query performance, they inevitably introduce additional space overhead. 
On the other hand, PLA-based learned compressors eliminate the need for such auxiliary structures. 
As shown in Figure~\ref{subfig:segment_skip}, each segment in a learned compressor naturally encodes the possible range of its covered keys, thus effectively performing the same function as skip pointers without incurring extra space costs.


\subsection{KV Store Compression}
Key-value (KV) storage is a popular DBMS paradigm where key-value pairs are treated as first-class citizens.  
The underlying storage layer of modern KV systems is the log-structured merge tree (LSM-tree), which buffers all writes into an in-memory structure (MemTable) before flushing them to files on disk (SSTable) and merging them through sequential I/Os~\cite{o1996log,luo2020lsm,huynh2024towards}. 
This design offers several key advantages, including enhanced write performance, efficient space utilization, and simplified concurrency control and recovery mechanisms, making KV systems particularly suitable for caching, session management, and real-time analytics. 

In practical KV systems like Redis~\cite{redis}, LevelDB~\cite{leveldb}, and RocksDB~\cite{rocksdb}, the key blocks and data blocks within an SSTable are often compressed to further reduce I/Os\footnote{\url{https://github.com/facebook/rocksdb/wiki/Compression}}. 
Current systems commonly employ generic byte-stream compressors, such as LZ4~\cite{lz4} and Snappy~\cite{snappy}, which are known to be more suitable for textual data rather than integer lists. 

As highlighted in Section~\ref{subsec:benchmark}, learned compressors offer superior compression efficacy and decompression throughput, making them a compelling alternative to conventional methods used for decades. 
For example, LevelDB~\cite{leveldb} uses Snappy~\cite{snappy} as the default compressor, with a decompression throughput of \textbf{546 MB/s}. 
In contrast, the SIMD-optimized learned compressor delivers a much better compression ratio and a \textbf{12.26\texttimes} higher throughput of \textbf{6.535 GiB/s}, according to our preliminary benchmark results in Section~\ref{subsec:benchmark}. 

Integrating learned compressors into practical KV stores requires efficient support for \emph{merge} operations, a.k.a.~the compaction operations in LSM-trees. 
A merge operation combines two compressed sorted key sets into a single one while maintaining the order. 
A key technical challenge here lies in dynamically adjusting the error constraint $\epsilon$ to accommodate the new data distribution resulting from the merging operation. 
Additionally, to boost read requests in KV systems, auxiliary index structures (e.g., compact B+-tree variants~\cite{singhal2001modern} or learned indexes~\cite{kraska2018case}) are necessary to enable fast localization of the corresponding line segment for a given key.


\subsection{DBMS Query Processing}
Learned compressors also hold substantial potential for traditional DBMS workloads, such as analytical query processing and large-scale distributed join acceleration. 

As introduced in Section~\ref{sec:pre}, a learned compressor encodes the inverse CDF (ICDF) of a given key set in an error-bounded and space-efficient manner. 
Thus, learned compressors can be directly applied to process \textsf{QUANTILE} queries and \textsf{MEDIAN} queries (a special case of \textsf{QUANTILE}), which are common yet costly operators in analytical workloads~\cite{chen2020survey}. 
For example, given a learned compressor $\mathcal{K}^c=(f, \Delta)$, the $k$-th $q$-quantile can be computed by $f(\lfloor N\cdot k/q\rfloor) + \Delta[\lfloor N\cdot k/q\rfloor]$. 
Based on the previous theoretical results~\cite{boffa2022learned,ferragina2020learned}, in Table~\ref{tab:complexity_quantile}, we summarize the space and time complexities for using learned compressors to compute arbitrary quantiles, contrasted with the commonly used quick select algorithm~\cite{hoare1961algorithm}. 

\begin{table}[h]
    \centering
    \begin{tabular}{|c|c|c|}
    \hline
        \textbf{\textsf{Method}} & \textbf{\textsf{Space Complexity}} & \textbf{\textsf{Time Complexity}} \\\hline
        \textsf{LC} (exact) & $O(N/\epsilon^2 + N\cdot\log\epsilon)$ & $O(\log(N/\epsilon^2))$\\
        \textsf{LC} (approx.) & $O(N/\epsilon^2)$ & $O(\log(N/\epsilon^2))$\\
        \textsf{QuickSel}~\cite{hoare1961algorithm} & $O(N)$ & Avg.~$O(N)$ \\\hline
    \end{tabular}
    \caption{Complexities to compute quantile queries. The logarithmic term in the time complexity comes from finding the corresponding segment using binary search.}
    \label{tab:complexity_quantile}
    \vspace{-1ex}
\end{table}

Notably, when query accuracy requirements can be relaxed, the residuals $\Delta$ for lossless reconstruction can be omitted, enabling approximate query processing (AQP) with a maximum error bounded by $\epsilon$. 
This adjustment yields a more compact representation consisting solely of \emph{segments}, offering flexibility to adapt learned compressors to the specific needs of analytical workloads. 
Additionally, it would be interesting to integrate the learned structures with conventional data summary techniques designed for AQP~\cite{chen2020survey}. 

Another promising direction for future exploration is the integration of learned compressors with join processing. 
In distributed environments, space-efficient filters like Bloom filters~\cite{guo2006theory,lang2019performance} are widely used to reduce communication costs, as exemplified by the BloomJoin technique~\cite{mackert1986r,li1995perf}. 
To compute $R_1\Join R_2$ where relations $R_1$ and $R_2$ reside on different sites, the BloomJoin technique involves three steps:
\ding{182} Site 1 creates a Bloom filter $BF(R_1)$ on the join attribute of $R_1$ and sends it to site 2. 
\ding{183} Site 2 filters $R_2$ by using $BF(R_1)$ and sends back tuples $R_{12}$ with matched join attribute to $R_1$. 
\ding{184} Finally, site 1 produces the final join result by joining $R_1$ with $R_{12}$ using local join algorithms like merge join. 

As a compact data representation, learned compressors can also function as effective data filters, akin to $BF(R_1)$ in BloomJoin. 
It would be interesting to benchmark the filtering efficacy of learned compressors under varying space constraints when compared to Bloom filters and their variants~\cite{guo2006theory,luo2018optimizing,liu2020stable}. 
A key technical challenge lies in the dependency of learned compressor performance on the key distribution characteristics, which can vary significantly for arbitrary SPJ (select-project-join) queries\footnote{Notably, this limitation is not unique to learned compressors but applies to all compression algorithms, as higher information entropy in a dataset inherently makes it harder to compress. }.
To mitigate the risk of inconsistent performance, a possible solution is to integrate learned compressors with conventional Bloom filters or recent learned Bloom filters~\cite{mitzenmacher2018model,liu2020stable}.
This hybrid approach leverages the strengths of both techniques to ensure robust and efficient query processing.

\begin{figure}
    \centering
    \includegraphics[width=0.45\textwidth]{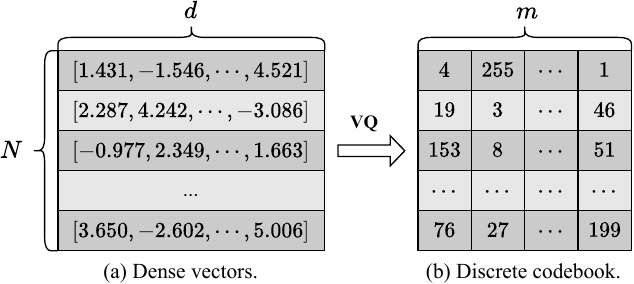}
    \caption{Illustration of a vector quantizer (VQ). When product quantization (PQ) is applied, dense vectors are partitioned into subspaces and clustered within each subspace. Each integer in the learned codebook corresponds to a cluster ID. }
    \label{fig:pq_eg}
    \vspace{-3ex}
\end{figure}

\subsection{Vector Database Compression}
The rapid advancements of retrieval-augmented generation (RAG) have recently sparked renewed interest in approximate nearest neighbor (ANN) search and VectorDB~\cite{lewis2020retrieval}. 
Due to the inefficiencies associated with storing and querying raw dense vectors in terms of both space and time, modern VectorDBs, such as Milvus\cite{milvus} and Pinecone~\cite{pinecone}, often employ coarse vector quantization (VQ) techniques, such as PQ~\cite{jegou2010product}, OPQ~\cite{ge2013optimized}, and AQ~\cite{babenko2014additive}. 
As illustrated in Figure~\ref{fig:pq_eg}, these methods represent dense vectors as discrete and compact representations (commonly referred to as codewords) while preserving the essential distance information. 

However, for web-scale vector sets, these compact codebooks remain too large, making it hard to deploy on resource-constrained devices, such as mobile phones or personal laptops. 
A recent study DeltaPQ~\cite{wang2020deltapq} attempts to compress VQ codebooks by exploring redundancies residing in codewords. 
Given the compression efficacy and high decompression throughput of learned compressors on integers, a promising approach is to reorder the codebook and apply learned compressors, potentially leading to a more compact vector representation while maintaining query efficiency. 
A promising research direction involves extending existing codebook traversal algorithms, such as PQFS~\cite{andre2016cache} and Quicker ADC~\cite{andre2016cache}, to accommodate compressed VQ codebooks. 
Furthermore, it would also be interesting to explore the potential of applying the ``learned'' methodology to compress other widely-used ANN indexes, such as graph-based indexes like HNSW~\cite{wang2021comprehensive} and locality-sensitive hashing (LSH) techniques~\cite{chi2017hashing}.

\subsection{Storage Mapping Table Compression}
Lastly, learned compressors also exhibit potential in low-level system optimizations. 
In main memory and secondary storage of computing systems~\cite{tang2020enhancing, zhou2017understanding, yan2019translation}, mapping table entries that map logical addresses to physical addresses can be intercepted and compressed into compression units, and each unit can hold multiple mapping table entries~\cite{pan2021hcftl}. 
The learned compressor aims to further improve compression and decompression efficiency, which can be naturally deployed in those existing approaches, for not only saving the space to hold the mapping table entries, but also speeding up address translation, since it requires fewer operations on fetching the entries and less time overhead for decompression.

\section{Technical Challenges and Future Directions}\label{sec:challenge}
Though learned compressors exhibit significant potential across various domains, they also present critical technical challenges that simultaneously serve as research opportunities for deeper exploration.

\subsection{Hyper-parameter Tuning}
As outlined in Section~\ref{sec:pre}, the error constraint $\epsilon$ is critical in determining the overall space cost of learned compressors. 
Assuming that all keys are \emph{i.i.d.}~samples from the same distribution, a recent work, BitTuner~\cite{bittuner}, derives a closed-form solution for minimizing the total space cost as $\epsilon^\mathsf{opt} = \sqrt{2\ln2\cdot\sigma^2\cdot C\cdot (K+2F)}$. 
However, the statistical assumption might be too strong in real applications, making a global hyper-parameter configuration sub-optimal. 

\begin{figure}[t]
    \centering
    \includegraphics[width=0.45\textwidth]{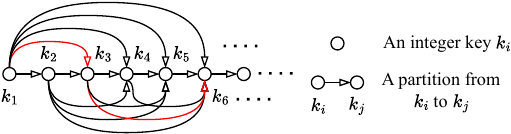}
    \caption{A graph model of the key partition problem. The path in red $k_1\to k_3 \to k_6$ implies partitioning $\{k_1,k_2,k_3,k_4,k_5,k_6\}$ into two parts $\{k_1,k_2\}$ and $\{k_3,k_4,k_5,k_6\}$.}
    \label{fig:partition}
    \vspace{-2ex}
\end{figure}

A possible solution is to adaptively partition the data into consecutive, disjoint partitions, with $\epsilon$ configured individually for each partition~\cite{boffa2021learned,boffa2022learned,ottaviano2014partitioned}. 
As illustrated in Figure~\ref{fig:partition}, such a data partition problem can be modeled as a graph $G=(V,E)$, where each node represents a key $k_i$, and each edge $k_i\to k_j$ ($j>i$) corresponds to a partition range $[i,j)$. 
By assigning edge weight based on the space cost of a learned compressor trained on the corresponding partition, the optimal partitioning can be obtained by finding the shortest path from $k_1$ to $k_N$. 
As $|V|=\Theta(N)$ and $|E|=\Theta(N^2)$, a standard Dijkstra's algorithm yields a time complexity of $O(N^2)$, which is intractable for large-scale applications. 
Though linear time approximations are investigated in~\cite{boffa2021learned,boffa2022learned,ottaviano2014partitioned}, their performance is still costly, given that fitting an optimal PLA only requires linear time~\cite{ORourke81}. 
Therefore, developing effective parameter tuning strategies that balance efficiency and effectiveness remains a challenging yet promising area of research.

\subsection{Handling Dynamic Updates}
A well-known technical challenge for learned indexes is handling dynamic operations, such as insertions, deletions, or modifications~\cite{ding2020alex,wongkham2022updatable}. 
The core difficulty lies in the need to promptly update outdated ML models to prevent performance degradation. 
Similarly, as learned compressors also incorporate ML models into their design, they inherit the same challenge of being difficult to update. 

Existing updatable learned index structures, such as ALEX~\cite{ding2020alex}, usually adopt a ``gapped array'' design, which reserves empty cells within the underlying key storage to enable efficient updates. 
However, this approach is incompatible with learned compressors, as the objective of compression is to represent data as compactly as possible, which fundamentally conflicts with the gapped array design. 
Moreover, when handling dynamic updates, not only does the model itself require updates, but the hyper-parameters of the learned compressor (e.g., the error constraint $\epsilon$), also need to be accordingly adjusted to accommodate data distribution shifts after updates. 

In the worst case, updating either the ML models (e.g., PLA) or the hyper-parameters may necessitate a complete reconstruction of the learned compressor, which can be computationally expensive. 
Consequently, devising effective update strategies is a critical technical challenge for deploying learned compressors in real-world systems with intensive data updates.

\subsection{Generic ML Model Choice}
Recall that the basic idea behind learned compressors is \emph{Data = Model Predictions + Residuals}. 
The ML model functions as a coarser predictor to reduce the residual range, such that each residual can be encoded using only a few bits. 
As introduced in Section~\ref{sec:pre}, existing learned compressors commonly adopt the error-bounded piecewise linear approximation ($\epsilon$-PLA) model. 
The reasons are mainly twofold. 
\ding{182}~Linear functions are simple to implement and computationally efficient compared to more complex models, particularly those relying on heavy deep learning frameworks like PyTorch. 
\ding{183}~Theoretical results from learned indexes using linear models, such as those in\cite{ferragina2020learned,liu2024learned}, can be extended to prove the effectiveness of learned compressors. 

Clearly, a design trade-off exists here. 
Simpler models, such as linear models, are easier to implement, well-supported by theoretical results, and benefit from a wide range of optimizations like~\cite{ORourke81}. 
However, more expressive models have the potential to achieve a better trade-off between fitting error and model complexity. 
It would be interesting to establish new theoretical results for non-linear base models (e.g., piecewise polynomial functions) and to develop efficient training and inference algorithms for such models\footnote{Intrinsically, the problem is to explore the learning theory on the hardness of fitting an ICDF within an arbitrary error constraint.}. 
Moreover, given the dual relationship between learned compression and learned indexing, new theoretical findings could also inspire the design of next-generation learned index structures.

\begin{figure}[t]
     \centering
     \begin{subfigure}[b]{0.4\textwidth}
         \centering
        \includegraphics[width=\textwidth]{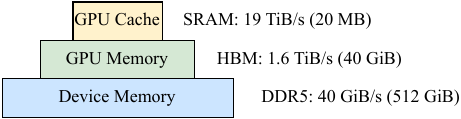}
        \caption{GPU memory hierarchy that highlights memory bandwidth and size.} 
        \label{fig:gpu_mem}
     \end{subfigure}
     
     \begin{subfigure}[b]{0.4\textwidth}
         \centering
         \includegraphics[width=\textwidth]{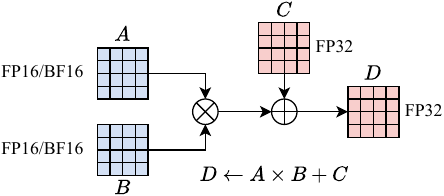}
         \caption{Tensor Cores on the NVIDIA A100 GPU. Each Tensor Core performs \textbf{64} mixed-precision floating-point fused multiply-add (FMA) operations \textbf{per cycle}.}
         \label{fig:tensor_core}
     \end{subfigure}
        \caption{Illustration of the GPU architecture characteristics. }
        \label{fig:gpu}
        \vspace{-3ex}
\end{figure}

\subsection{Extension to Floating-Point Data Compression}
Learned compressors are specifically designed for \emph{integer data}. 
However, in other domains, such as scientific computing and sensor data analytics, users often work with floating-point data of \emph{bounded precision}~\cite{liu2021decomposed}. 
For example, a typical GPS device generates data ranging from $-180.0000$ to $180.0000$, and a resistance thermometer records temperature data within the range of $-50.000 {^\circ\text{C}}$ to $100.000 {^\circ\text{C}}$. 
Existing solutions~\cite{pelkonen2015gorilla,liu2021decomposed,blalock2018sprintz} often employ quantization and dictionary encoding to compress large-scale floating-point values. 

According to the IEEE standard~\cite{8766229}, a floating value is defined as $(-1)^{\text{sign}}\times 1.{\text{mantissa}}\times 2^{\text{exponent}}$, where the number of bits allocated to the mantissa and exponent determines the precision. 
To adapt learned compressors for bounded floats, a straightforward approach is to separately compress the mantissa and exponent, each of which can be treated as an integer, using learned compressors. 
However, three technical questions here require further investigation. 
\ding{182}~Can this approach achieve comparable performance to compressing integer data?
\ding{183}~Scientific computing workloads often require \emph{in situ} query processing~\cite{chen2024fcbench}. How can direct query processing on compressed data be enabled to maintain compatibility with existing applications?
\ding{184}~If some level of floating-point precision can be sacrificed, could this trade-off result in significant gains in compression ratios and decompression performance?

\subsection{Acceleration using Modern Hardware}
Although initially designed as accelerators for rendering, GPUs are playing an increasingly important role in AI model training and high-performance computing. 
Due to the highly multi-threaded architecture, GPUs can accelerate a wide range of classic data-centric tasks, such as sorting~\cite{govindaraju2006gputerasort, rui2020efficient}, query processing~\cite{paul2020improving}, and indexing~\cite{awad2023analyzing,heo2020iiu}. 
Given that our SIMD-optimized learned compressor already outperforms CPU baselines, extending it to the GPU's single-instruction multiple-thread (SIMT) operating model presents a promising avenue for further performance enhancement. 

To align with the characteristics of GPU architectures (as illustrated in Figure~\ref{fig:gpu}), we outline key optimizations to be explored as follows. 
\ding{182}~To fully leverage the GPU's highly multi-threaded design, it is necessary to further reduce the data dependencies inherent in learned compressors to maximize parallelism. 
\ding{183}~Similar to the SIMD case, memory access latency constitutes a significant portion of the total computational overhead. 
Refining the learned compressor's structure to better align with the GPU memory hierarchy is essential for achieving better performance. 
\ding{184}~Modern GPUs like NVIDIA A100 are equipped with Tensor Cores capable of operating 64 fused multiply-add (FMA, i.e., $D\gets A\times B + C$) operations per clock. 
This hardware capability is well-suited to PLA-based learned compressors, given that key recovery is to compute $\lfloor \alpha\cdot(i-s) + \beta \rfloor + \Delta[i]$. 
However, as depicted in Figure~\ref{fig:tensor_core}, the Tensor Cores are designed for \emph{mixed-precision} AI model training, where $A$, $B$ are matrices of FP16/BP16 floats while $C$, $D$ are matrices of FP32 floats. 
Therefore, addressing the numerical issues introduced by mixed-precision computation will emerge as a technical challenge for future research. 

%

\section{Conclusion}\label{sec:conclusion}
Compressing integer keys is fundamental across diverse scenarios, spanning from low-level system optimizations to high-level data-driven applications. 
While research on learned compressors is still in its infancy, our preliminary benchmarks highlight that an SIMD-optimized learned compressor can significantly outperform conventional data compression algorithms. 
Building on these findings, this paper envisions a future where learned compression techniques become vital components of modern database and information retrieval systems. 
However, despite their significant potential, learned compressors face critical technical challenges that also present compelling research opportunities, paving the way for deeper exploration and innovation in this emerging field.


\bibliographystyle{IEEEtran}
\bibliography{ref}

\begin{thebibliography}{10}
\providecommand{\url}[1]{#1}
\csname url@samestyle\endcsname
\providecommand{\newblock}{\relax}
\providecommand{\bibinfo}[2]{#2}
\providecommand{\BIBentrySTDinterwordspacing}{\spaceskip=0pt\relax}
\providecommand{\BIBentryALTinterwordstretchfactor}{4}
\providecommand{\BIBentryALTinterwordspacing}{\spaceskip=\fontdimen2\font plus
\BIBentryALTinterwordstretchfactor\fontdimen3\font minus
  \fontdimen4\font\relax}
\providecommand{\BIBforeignlanguage}[2]{{%
\expandafter\ifx\csname l@#1\endcsname\relax
\typeout{** WARNING: IEEEtran.bst: No hyphenation pattern has been}%
\typeout{** loaded for the language `#1'. Using the pattern for}%
\typeout{** the default language instead.}%
\else
\language=\csname l@#1\endcsname
\fi
#2}}
\providecommand{\BIBdecl}{\relax}
\BIBdecl

\bibitem{marx2013big}
V.~Marx, ``The big challenges of big data,'' \emph{Nature}, vol. 498, no. 7453,
  pp. 255--260, 2013.

\bibitem{trelles2011big}
O.~Trelles, P.~Prins, M.~Snir, and R.~C. Jansen, ``Big data, but are we
  ready?'' \emph{Nature Reviews Genetics}, vol.~12, no.~3, pp. 224--224, 2011.

\bibitem{satyanarayanan2011mobile}
M.~Satyanarayanan, ``Mobile computing: the next decade,'' \emph{ACM SIGMOBILE
  Mobile Computing and Communications Review}, vol.~15, no.~2, pp. 2--10, 2011.

\bibitem{lin2019computation}
L.~Lin, X.~Liao, H.~Jin, and P.~Li, ``Computation offloading toward edge
  computing,'' \emph{Proceedings of the IEEE}, vol. 107, no.~8, pp. 1584--1607,
  2019.

\bibitem{rocksdb}
``{RocksDB},'' \url{https://rocksdb.org/}, accessed: 2024-10-20.

\bibitem{luo2020lsm}
C.~Luo and M.~J. Carey, ``Lsm-based storage techniques: a survey,'' \emph{The
  VLDB Journal}, vol.~29, no.~1, pp. 393--418, 2020.

\bibitem{ozsu1999principles}
M.~T. {\"O}zsu, P.~Valduriez \emph{et~al.}, \emph{Principles of distributed
  database systems}.\hskip 1em plus 0.5em minus 0.4em\relax Springer, 1999,
  vol.~2.

\bibitem{geambasu2010comet}
R.~Geambasu, A.~A. Levy, T.~Kohno, A.~Krishnamurthy, and H.~M. Levy, ``Comet:
  An active distributed $\{$Key-Value$\}$ store,'' in \emph{9th USENIX
  Symposium on Operating Systems Design and Implementation (OSDI 10)}, 2010.

\bibitem{pibiri2020techniques}
G.~E. Pibiri and R.~Venturini, ``Techniques for inverted index compression,''
  \emph{ACM Computing Surveys (CSUR)}, vol.~53, no.~6, pp. 1--36, 2020.

\bibitem{pan2024survey}
J.~J. Pan, J.~Wang, and G.~Li, ``Survey of vector database management
  systems,'' \emph{The VLDB Journal}, vol.~33, no.~5, pp. 1591--1615, 2024.

\bibitem{elias1975universal}
P.~Elias, ``Universal codeword sets and representations of the integers,''
  \emph{IEEE transactions on information theory}, vol.~21, no.~2, pp. 194--203,
  1975.

\bibitem{golomb1966run}
S.~Golomb, ``Run-length encodings (corresp.),'' \emph{IEEE transactions on
  information theory}, vol.~12, no.~3, pp. 399--401, 1966.

\bibitem{teuhola1978compression}
J.~Teuhola, ``A compression method for clustered bit-vectors,''
  \emph{Information processing letters}, vol.~7, no.~6, pp. 308--311, 1978.

\bibitem{rice1971adaptive}
R.~Rice and J.~Plaunt, ``Adaptive variable-length coding for efficient
  compression of spacecraft television data,'' \emph{IEEE Transactions on
  Communication Technology}, vol.~19, no.~6, pp. 889--897, 1971.

\bibitem{stepanov2011simd}
A.~A. Stepanov, A.~R. Gangolli, D.~E. Rose, R.~J. Ernst, and P.~S. Oberoi,
  ``Simd-based decoding of posting lists,'' in \emph{Proceedings of the 20th
  ACM international conference on Information and knowledge management}, 2011,
  pp. 317--326.

\bibitem{dean2009challenges}
J.~Dean \emph{et~al.}, ``Challenges in building large-scale information
  retrieval systems,'' in \emph{Keynote of the 2nd ACM international conference
  on web search and data mining (WSDM)}, vol.~10, no. 1498759.1498761, 2009.

\bibitem{zukowski2006super}
M.~Zukowski, S.~Heman, N.~Nes, and P.~Boncz, ``Super-scalar ram-cpu cache
  compression,'' in \emph{22nd International Conference on Data Engineering
  (ICDE'06)}.\hskip 1em plus 0.5em minus 0.4em\relax IEEE, 2006, pp. 59--59.

\bibitem{yan2009inverted}
H.~Yan, S.~Ding, and T.~Suel, ``Inverted index compression and query processing
  with optimized document ordering,'' in \emph{Proceedings of the 18th
  international conference on World wide web}, 2009, pp. 401--410.

\bibitem{elias1974efficient}
P.~Elias, ``Efficient storage and retrieval by content and address of static
  files,'' \emph{Journal of the ACM (JACM)}, vol.~21, no.~2, pp. 246--260,
  1974.

\bibitem{ottaviano2014partitioned}
G.~Ottaviano and R.~Venturini, ``Partitioned elias-fano indexes,'' in
  \emph{Proceedings of the 37th international ACM SIGIR conference on Research
  \& development in information retrieval}, 2014, pp. 273--282.

\bibitem{vigna2013quasi}
S.~Vigna, ``Quasi-succinct indices,'' in \emph{Proceedings of the sixth ACM
  international conference on Web search and data mining}, 2013, pp. 83--92.

\bibitem{moffat1996exploiting}
A.~Moffat and L.~Stuiver, ``Exploiting clustering in inverted file
  compression,'' in \emph{Proceedings of Data Compression
  Conference-DCC'96}.\hskip 1em plus 0.5em minus 0.4em\relax IEEE, 1996, pp.
  82--91.

\bibitem{moffat2000binary}
------, ``Binary interpolative coding for effective index compression,''
  \emph{Information Retrieval}, vol.~3, pp. 25--47, 2000.

\bibitem{huffman1952method}
D.~A. Huffman, ``A method for the construction of minimum-redundancy codes,''
  \emph{Proceedings of the IRE}, vol.~40, no.~9, pp. 1098--1101, 1952.

\bibitem{duda2013asymmetric}
J.~Duda, ``Asymmetric numeral systems: entropy coding combining speed of
  huffman coding with compression rate of arithmetic coding,'' \emph{arXiv
  preprint arXiv:1311.2540}, 2013.

\bibitem{moffat1998arithmetic}
A.~Moffat, R.~M. Neal, and I.~H. Witten, ``Arithmetic coding revisited,''
  \emph{ACM Transactions on Information Systems (TOIS)}, vol.~16, no.~3, pp.
  256--294, 1998.

\bibitem{kraska2018case}
T.~Kraska, A.~Beutel, E.~H. Chi, J.~Dean, and N.~Polyzotis, ``The case for
  learned index structures,'' in \emph{Proceedings of the 2018 international
  conference on management of data}, 2018, pp. 489--504.

\bibitem{boffa2022learned}
A.~Boffa, P.~Ferragina, and G.~Vinciguerra, ``A learned approach to design
  compressed rank/select data structures,'' \emph{ACM Transactions on
  Algorithms (TALG)}, vol.~18, no.~3, pp. 1--28, 2022.

\bibitem{boffa2021learned}
------, ``A ``learned'' approach to quicken and compress rank/select
  dictionaries,'' in \emph{2021 Proceedings of the Workshop on Algorithm
  Engineering and Experiments (ALENEX)}.\hskip 1em plus 0.5em minus 0.4em\relax
  SIAM, 2021, pp. 46--59.

\bibitem{ferragina2020pgm}
P.~Ferragina and G.~Vinciguerra, ``The pgm-index: a fully-dynamic compressed
  learned index with provable worst-case bounds,'' \emph{Proceedings of the
  VLDB Endowment}, vol.~13, no.~8, pp. 1162--1175, 2020.

\bibitem{ferragina2020learned}
P.~Ferragina, F.~Lillo, and G.~Vinciguerra, ``Why are learned indexes so
  effective?'' in \emph{International Conference on Machine Learning}.\hskip
  1em plus 0.5em minus 0.4em\relax PMLR, 2020, pp. 3123--3132.

\bibitem{postgres}
``{PostgreSQL},'' \url{https://www.postgresql.org/}, accessed: 2024-10-20.

\bibitem{oracle}
``{Oracle},'' \url{https://www.oracle.com/}, accessed: 2024-10-20.

\bibitem{solr}
``{Apache Solr},'' \url{https://solr.apache.org/}, accessed: 2024-10-20.

\bibitem{elasticsearch}
``{Elasticsearch: The Official Distributed Search \& Analytics Engine},''
  \url{https://www.elastic.co/elasticsearch}, accessed: 2024-10-20.

\bibitem{ccnews2020}
J.~Mackenzie, R.~Benham, M.~Petri, J.~R. Trippas, J.~S. Culpepper, and
  A.~Moffat, ``Cc-news-en: A large english news corpus,'' in \emph{Proc. CIKM},
  2020, pp. 3077--3084.

\bibitem{chaudhuri2017approximate}
S.~Chaudhuri, B.~Ding, and S.~Kandula, ``Approximate query processing: No
  silver bullet,'' in \emph{Proceedings of the 2017 ACM International
  Conference on Management of Data}, 2017, pp. 511--519.

\bibitem{cormode2011sketch}
G.~Cormode, ``Sketch techniques for approximate query processing,''
  \emph{Foundations and Trends in Databases. NOW publishers}, vol.~15, 2011.

\bibitem{ORourke81}
J.~O'Rourke, ``An on-line algorithm for fitting straight lines between data
  ranges,'' \emph{Commun. {ACM}}, vol.~24, no.~9, pp. 574--578, 1981.

\bibitem{liu2024learned}
Q.~Liu, S.~Han, Y.~Qi, J.~Peng, J.~Li, L.~Lin, and L.~Chen, ``Why are learned
  indexes so effective but sometimes ineffective?'' \emph{arXiv preprint
  arXiv:2410.00846}, 2024.

\bibitem{bittuner}
``{BitTuner},'' \url{https://github.com/qyliu-hkust/BitTuner}, accessed:
  2024-10-20.

\bibitem{ding2020alex}
J.~Ding, U.~F. Minhas, J.~Yu, C.~Wang, J.~Do, Y.~Li, H.~Zhang, B.~Chandramouli,
  J.~Gehrke, D.~Kossmann \emph{et~al.}, ``Alex: an updatable adaptive learned
  index,'' in \emph{Proceedings of the 2020 ACM SIGMOD International Conference
  on Management of Data}, 2020, pp. 969--984.

\bibitem{serfling2009approximation}
R.~J. Serfling, \emph{Approximation theorems of mathematical statistics}.\hskip
  1em plus 0.5em minus 0.4em\relax John Wiley \& Sons, 2009.

\bibitem{graefe2011modern}
G.~Graefe \emph{et~al.}, ``Modern b-tree techniques,'' \emph{Foundations and
  Trends{\textregistered} in Databases}, vol.~3, no.~4, pp. 203--402, 2011.

\bibitem{salad}
``{SALAD: SIMD-Aware Learned Data Compression},''
  \url{https://github.com/qyliu-hkust/salad/tree/PGM-INDEX-SIMD-TEMP},
  accessed: 2024-10-20.

\bibitem{trotman2016vacuo}
A.~Trotman and J.~Lin, ``In vacuo and in situ evaluation of simd codecs,'' in
  \emph{Proceedings of the 21st Australasian Document Computing Symposium},
  2016, pp. 1--8.

\bibitem{lee2012prefetching}
J.~Lee, H.~Kim, and R.~Vuduc, ``When prefetching works, when it doesn’t, and
  why,'' \emph{ACM Transactions on Architecture and Code Optimization (TACO)},
  vol.~9, no.~1, pp. 1--29, 2012.

\bibitem{singhal2001modern}
A.~Singhal \emph{et~al.}, ``Modern information retrieval: A brief overview,''
  \emph{IEEE Data Eng. Bull.}, vol.~24, no.~4, pp. 35--43, 2001.

\bibitem{o1996log}
P.~O’Neil, E.~Cheng, D.~Gawlick, and E.~O’Neil, ``The log-structured
  merge-tree (lsm-tree),'' \emph{Acta Informatica}, vol.~33, pp. 351--385,
  1996.

\bibitem{huynh2024towards}
A.~Huynh, H.~A. Chaudhari, E.~Terzi, and M.~Athanassoulis, ``Towards
  flexibility and robustness of lsm trees,'' \emph{The VLDB Journal}, pp.
  1--24, 2024.

\bibitem{redis}
``{Redis},'' \url{https://redis.io/}, accessed: 2024-10-20.

\bibitem{leveldb}
``{LevelDB},'' \url{https://github.com/google/leveldb}, accessed: 2024-10-20.

\bibitem{lz4}
``{Etremely fast compression},'' \url{https://lz4.org/}, accessed: 2024-06-12.

\bibitem{snappy}
``{Snappy: A fast compressor/decompressor},''
  \url{https://google.github.io/snappy/}, accessed: 2024-06-12.

\bibitem{chen2020survey}
Z.~Chen and A.~Zhang, ``A survey of approximate quantile computation on
  large-scale data,'' \emph{IEEE Access}, vol.~8, pp. 34\,585--34\,597, 2020.

\bibitem{hoare1961algorithm}
C.~A. Hoare, ``Algorithm 65: find,'' \emph{Communications of the ACM}, vol.~4,
  no.~7, pp. 321--322, 1961.

\bibitem{guo2006theory}
D.~Guo, J.~Wu, H.~Chen, and X.~Luo, ``Theory and network applications of
  dynamic bloom filters,'' in \emph{Proceedings IEEE INFOCOM 2006. 25TH IEEE
  International Conference on Computer Communications}.\hskip 1em plus 0.5em
  minus 0.4em\relax Citeseer, 2006, pp. 1--12.

\bibitem{lang2019performance}
H.~Lang, T.~Neumann, A.~Kemper, and P.~Boncz, ``Performance-optimal filtering:
  Bloom overtakes cuckoo at high throughput,'' \emph{Proceedings of the VLDB
  Endowment}, vol.~12, no.~5, pp. 502--515, 2019.

\bibitem{mackert1986r}
L.~F. Mackert and G.~M. Lohman, ``R* optimizer validation and performance
  evaluation for local queries,'' in \emph{Proceedings of the 1986 ACM SIGMOD
  international conference on Management of data}, 1986, pp. 84--95.

\bibitem{li1995perf}
Z.~Li and K.~A. Ross, ``Perf join: An alternative to two-way semijoin and
  bloomjoin,'' in \emph{Proceedings of the fourth international conference on
  Information and knowledge management}, 1995, pp. 137--144.

\bibitem{luo2018optimizing}
L.~Luo, D.~Guo, R.~T. Ma, O.~Rottenstreich, and X.~Luo, ``Optimizing bloom
  filter: Challenges, solutions, and comparisons,'' \emph{IEEE Communications
  Surveys \& Tutorials}, vol.~21, no.~2, pp. 1912--1949, 2018.

\bibitem{liu2020stable}
Q.~Liu, L.~Zheng, Y.~Shen, and L.~Chen, ``Stable learned bloom filters for data
  streams,'' \emph{Proceedings of the VLDB Endowment}, vol.~13, no.~12, pp.
  2355--2367, 2020.

\bibitem{mitzenmacher2018model}
M.~Mitzenmacher, ``A model for learned bloom filters and optimizing by
  sandwiching,'' \emph{Advances in Neural Information Processing Systems},
  vol.~31, 2018.

\bibitem{lewis2020retrieval}
P.~Lewis, E.~Perez, A.~Piktus, F.~Petroni, V.~Karpukhin, N.~Goyal,
  H.~K{\"u}ttler, M.~Lewis, W.-t. Yih, T.~Rockt{\"a}schel \emph{et~al.},
  ``Retrieval-augmented generation for knowledge-intensive nlp tasks,''
  \emph{Advances in Neural Information Processing Systems}, vol.~33, pp.
  9459--9474, 2020.

\bibitem{milvus}
``{The high-performance vector database built for scale},''
  \url{https://milvus.io/}, accessed: 2024-06-12.

\bibitem{pinecone}
``{Pinecone: The vector database to build knowledgeable AI},''
  \url{https://www.pinecone.io/}, accessed: 2024-06-12.

\bibitem{jegou2010product}
H.~Jegou, M.~Douze, and C.~Schmid, ``Product quantization for nearest neighbor
  search,'' \emph{IEEE transactions on pattern analysis and machine
  intelligence}, vol.~33, no.~1, pp. 117--128, 2010.

\bibitem{ge2013optimized}
T.~Ge, K.~He, Q.~Ke, and J.~Sun, ``Optimized product quantization,'' \emph{IEEE
  transactions on pattern analysis and machine intelligence}, vol.~36, no.~4,
  pp. 744--755, 2013.

\bibitem{babenko2014additive}
A.~Babenko and V.~Lempitsky, ``Additive quantization for extreme vector
  compression,'' in \emph{Proceedings of the IEEE Conference on Computer Vision
  and Pattern Recognition}, 2014, pp. 931--938.

\bibitem{wang2020deltapq}
R.~Wang and D.~Deng, ``Deltapq: lossless product quantization code compression
  for high dimensional similarity search,'' \emph{Proceedings of the VLDB
  Endowment}, vol.~13, no.~13, pp. 3603--3616, 2020.

\bibitem{andre2016cache}
F.~Andr{\'e}, A.-M. Kermarrec, and N.~Le~Scouarnec, ``Cache locality is not
  enough: High-performance nearest neighbor search with product quantization
  fast scan,'' in \emph{42nd International Conference on Very Large Data
  Bases}, vol.~9, no.~4, 2016, p.~12.

\bibitem{wang2021comprehensive}
M.~Wang, X.~Xu, Q.~Yue, and Y.~Wang, ``A comprehensive survey and experimental
  comparison of graph-based approximate nearest neighbor search,''
  \emph{Proceedings of the VLDB Endowment}, vol.~14, no.~11, pp. 1964--1978,
  2021.

\bibitem{chi2017hashing}
L.~Chi and X.~Zhu, ``Hashing techniques: A survey and taxonomy,'' \emph{ACM
  Computing Surveys (Csur)}, vol.~50, no.~1, pp. 1--36, 2017.

\bibitem{tang2020enhancing}
X.~Tang, Z.~Zhang, W.~Xu, M.~T. Kandemir, R.~Melhem, and J.~Yang, ``Enhancing
  address translations in throughput processors via compression,'' in
  \emph{Proceedings of the ACM International Conference on Parallel
  Architectures and Compilation Techniques}, 2020, pp. 191--204.

\bibitem{zhou2017understanding}
Y.~Zhou, F.~Wu, P.~Huang, X.~He, C.~Xie, and J.~Zhou, ``Understanding and
  alleviating the impact of the flash address translation on solid state
  devices,'' \emph{ACM Transactions on Storage (TOS)}, vol.~13, no.~2, pp.
  1--29, 2017.

\bibitem{yan2019translation}
Z.~Yan, D.~Lustig, D.~Nellans, and A.~Bhattacharjee, ``Translation ranger:
  Operating system support for contiguity-aware tlbs,'' in \emph{Proceedings of
  the 46th International Symposium on Computer Architecture}, 2019, pp.
  698--710.

\bibitem{pan2021hcftl}
Y.~Pan, H.~Chen, J.~Zhao, and Y.~Xu, ``Hcftl: A locality-aware flash
  translation layer for efficient address translation,'' \emph{IEEE
  Transactions on Computer-Aided Design of Integrated Circuits and Systems},
  vol.~41, no.~8, pp. 2477--2489, 2021.

\bibitem{wongkham2022updatable}
C.~Wongkham, B.~Lu, C.~Liu, Z.~Zhong, E.~Lo, and T.~Wang, ``Are updatable
  learned indexes ready?'' \emph{Proceedings of the VLDB Endowment}, vol.~15,
  no.~11, pp. 3004--3017, 2022.

\bibitem{liu2021decomposed}
C.~Liu, H.~Jiang, J.~Paparrizos, and A.~J. Elmore, ``Decomposed bounded floats
  for fast compression and queries,'' \emph{Proceedings of the VLDB Endowment},
  vol.~14, no.~11, pp. 2586--2598, 2021.

\bibitem{pelkonen2015gorilla}
T.~Pelkonen, S.~Franklin, J.~Teller, P.~Cavallaro, Q.~Huang, J.~Meza, and
  K.~Veeraraghavan, ``Gorilla: A fast, scalable, in-memory time series
  database,'' \emph{Proceedings of the VLDB Endowment}, vol.~8, no.~12, pp.
  1816--1827, 2015.

\bibitem{blalock2018sprintz}
D.~Blalock, S.~Madden, and J.~Guttag, ``Sprintz: Time series compression for
  the internet of things,'' \emph{Proceedings of the ACM on Interactive,
  Mobile, Wearable and Ubiquitous Technologies}, vol.~2, no.~3, pp. 1--23,
  2018.

\bibitem{8766229}
``Ieee standard for floating-point arithmetic,'' \emph{IEEE Std 754-2019
  (Revision of IEEE 754-2008)}, pp. 1--84, 2019.

\bibitem{chen2024fcbench}
X.~Chen, J.~Tian, I.~Beaver, C.~Freeman, Y.~Yan, J.~Wang, and D.~Tao,
  ``Fcbench: Cross-domain benchmarking of lossless compression for
  floating-point data,'' \emph{Proceedings of the VLDB Endowment}, vol.~17,
  no.~6, pp. 1418--1431, 2024.

\bibitem{govindaraju2006gputerasort}
N.~Govindaraju, J.~Gray, R.~Kumar, and D.~Manocha, ``Gputerasort: high
  performance graphics co-processor sorting for large database management,'' in
  \emph{Proceedings of the 2006 ACM SIGMOD international conference on
  Management of data}, 2006, pp. 325--336.

\bibitem{rui2020efficient}
R.~Rui, H.~Li, and Y.-C. Tu, ``Efficient join algorithms for large database
  tables in a multi-gpu environment,'' in \emph{Proceedings of the VLDB
  Endowment. International Conference on Very Large Data Bases}, vol.~14,
  no.~4.\hskip 1em plus 0.5em minus 0.4em\relax NIH Public Access, 2020, p.
  708.

\bibitem{paul2020improving}
J.~Paul, B.~He, S.~Lu, and C.~T. Lau, ``Improving execution efficiency of
  just-in-time compilation based query processing on gpus,'' \emph{Proceedings
  of the VLDB Endowment}, vol.~14, no.~2, pp. 202--214, 2020.

\bibitem{awad2023analyzing}
M.~A. Awad, S.~Ashkiani, S.~D. Porumbescu, M.~Farach-Colton, and J.~D. Owens,
  ``Analyzing and implementing gpu hash tables,'' in \emph{2023 Symposium on
  Algorithmic Principles of Computer Systems (APOCS)}.\hskip 1em plus 0.5em
  minus 0.4em\relax SIAM, 2023, pp. 33--50.

\bibitem{heo2020iiu}
J.~Heo, J.~Won, Y.~Lee, S.~Bharuka, J.~Jang, T.~J. Ham, and J.~W. Lee, ``Iiu:
  Specialized architecture for inverted index search,'' in \emph{Proceedings of
  the Twenty-Fifth International Conference on Architectural Support for
  Programming Languages and Operating Systems}, 2020, pp. 1233--1245.

\end{thebibliography}

\end{document}